\documentclass[runningheads]{llncs}
\usepackage[T1]{fontenc}
\usepackage{graphicx}
\usepackage{tabularx}
\usepackage{etoolbox}
\AfterEndEnvironment{table}{\vspace{-3mm}}  
\usepackage{adjustbox}
\usepackage{amsmath}
\usepackage[colorlinks=true,citecolor=blue,linkcolor=blue,urlcolor=blue]{hyperref}
\usepackage{float}
\usepackage{makecell}
\usepackage{microtype}
\begin{document}
\title{NIMARC-MRI: Abdominal HASTE Dataset and a Baseline U-Net Exposing the Synthetic-to-Real Gap in Low-Resource Motion Correction\thanks{Accepted for publication at the MIRASOL Workshop at MICCAI 2026.}}
%
%
%
\titlerunning{NIMARC-MRI: Abdominal MRI Dataset for Motion Correction}
\author{Abdulrazaq A. Zubair\inst{1,2}\orcidID{0009-0007-8010-3999} \and
Nafiu Musa Muhammad\inst{3,4}\orcidID{0009-0008-6224-1551} \and
Simeon Krah\inst{2}\orcidID{0009-0000-7566-1231} \and
Ummasalma Usman Ibrahim\inst{3,4}\orcidID{0009-0005-6929-3520} \and
Yusuf Tijjani Abdumumin\inst{4}\orcidID{0009-0006-9837-4968} \and
Ismail Ismail Tijjani\inst{4}\orcidID{0009-0003-3571-4214} \and
Ummulkhairi Ibrahim\inst{2,4}\orcidID{ 0009-0005-9542-8011} \and
Alyasaa Anas\inst{2,6}\orcidID{0009-0001-0797-118X} \and
Mubaraq Yakubu\inst{7}\orcidID{0009-0003-3394-5641} \and
Abbas Rabiu Muhammad\inst{3}\orcidID{0000-0002-7139-9341}}

\authorrunning{Zubair et al.}

\institute{Radiography and Radiation Sciences, Federal University of Health Sciences, Azare, (FUHSA) Nigeria\\
\email{abdulrazaq.zubair@fuhsa.edu.ng} \and
African Institute for Research Advancement and Innovation (AIRA Africa) \and
Bayero University Kano, Nigeria \and
Medserve Kano Diagnostic Center, Kano, Nigeria \and Northwest University kano, Nigeria \and
Radiology, University of Maiduguri Teaching Hospital, Maiduguri, Nigeria \and
PET Imaging Centre, King's College London, University of London, UK}
%
%
\maketitle              

\begin{abstract}
Respiratory motion degrades abdominal T2 HASTE MRI in low- and middle-income countries where vendor motion-correction licenses are unavailable and failed scans are deleted during routine PACS cleanup, precluding supervised training. To address this, we introduce NIMARC-MRI, the first public abdominal MRI dataset from West Africa, comprising 139 clean HASTE, 25 native motion-degraded HASTE, and 79 paired HASTE–TSE TRIGGER acquisitions from a Nigerian centre operating 1.5 T Siemens scanner without integrated motion-correction licenses. A 2D U-Net with 7.7 million parameters was trained on 3D-consistent synthetic respiratory motion necessitated by local infrastructure reality and validated via a three-tier strategy: held-out synthetic data, blinded radiologist Likert scoring on native real motion, and cross-sequence TRIGGER generalisation. On synthetic test data the model achieved SSIM 0.863 ± 0.042 and PSNR 30.06 ± 1.80 dB. On native real motion, however, blinded radiologist and radiographer scores showed no significant improvement (mean Likert 4.06 ± 0.55 original versus 4.04 ± 0.61 corrected, P = 0.914), with 28\% of cases rated worse after correction. Cross-sequence TRIGGER evaluation showed modest SSIM improvement (0.404 ± 0.073 versus 0.334 ± 0.055, P < 0.001) without radiologist-perceived gain. These findings demonstrate that conservative synthetic motion fails to capture clinical motion severity, exposing a reproducible synthetic-to-real gap. NIMARC-MRI is released on Zenodo (\url{https://doi.org/10.5281/zenodo.22115936}) under a controlled data-use agreement requiring citation. Sequence metadata and a sample subset are publicly accessible to facilitate discovery, while patient-level data remain restricted to approved collaborators. The aim is to establish a reproducible baseline for motion correction in resource-constrained settings.

\end{abstract}
\section{Introduction}
Respiratory motion degrades abdominal T2 half-Fourier acquisition single-shot turbo spin-echo (HASTE) MRI, where sub-second acquisition is vulnerable to breath-hold failure and diaphragmatic drift \cite{ref1}. In high-resource centers, vendor integrated motion-correction (PROPELLER, BLADE, MultiVane) oversamples central k-space to retrospectively correct corrupted blades \cite{ref2}. These techniques require sequence licenses and higher-field platforms frequently unavailable in low and middle-income countries (LMICs) \cite{ref3}. Across West Africa, MRI availability averages 0.30-0.48 units per million population \cite{ref4}, most are low-field systems, and high-field 1.5 T scanners are concentrated in private urban centers \cite{ref5}. Infrastructural limitations, unreliable power, helium scarcity, and absent vendor contracts constrain deployment of advanced motion-correction software \cite{ref1,ref6}. Most Nigerian tertiary sites, 1.5 T scanners lack motion-correction licenses; repeat scans cost a lot compared to the monthly income of the average citizen, and severely degraded acquisitions are deleted during routine PACS cleanup. Consequently, suboptimal images are often accepted for reporting, degrading diagnostic confidence \cite{ref6}.

Deep learning offers a vendor-neutral alternative. Image-domain U-Nets operate on reconstructed DICOM images without raw k-space access, making them deployable without vendor research agreements \cite{ref1,ref10}. However, k-space methods (MoPED, D2MC-Net) require raw data and vendor agreements inaccessible in most LMICs \cite{ref11,ref12}. The CUPID framework recently demonstrated physics driven reconstruction from DICOMs alone, reinforcing image-domain processing as the practical LMIC path \cite{ref13}. Supervised training requires paired motion corrupted inputs and clean targets. In our setting, failed scans are deleted during PACS cleanup, erasing the pairs required for supervised learning \cite{ref8}. Synthetic motion simulation becomes a necessity, not a convenience when real failed scans are unavailable \cite{ref14}. Domain-shift concerns motivate careful cross-sequence validation \cite{ref15,ref16}.

We introduce NIMARC-MRI, a dataset of native motion-degraded HASTE and paired HASTE--TSE-TRIGGER acquisitions from Northern Nigeria with full sequence characterization and ethics oversight. We train a baseline 2D U-Net on 3D-consistent synthetic respiratory motion and validate it via a three-tier strategy: held-out synthetic data, blinded radiologist scoring on native real motion, and cross-sequence TRIGGER generalisation. NIMARC-MRI is released on Zenodo (\url{https://doi.org/10.5281/zenodo.22115936}) under a controlled data-use agreement (National Health Research Ethics Committee approval: NHREC/17/03/2018: SHREC/2026/7923) permitting multi-project use by approved researchers; sequence metadata and a sample subset are publicly accessible to facilitate discovery, while patient-level data remain restricted to approved collaborators. The aim is to develop and validate a vendor-neutral image-domain deep learning pipeline for respiratory motion correction in abdominal HASTE MRI, and to establish NIMARC-MRI as a reproducible benchmark for abdominal MRI-AI research in West African populations.


\section{Previous Work}
\subsubsection{2.1 Image-domain deep learning for abdominal motion correction:} 
Image-to-image translation networks dominate retrospective motion correction because they operate on reconstructed DICOM images without raw k-space access \cite{ref7,ref8,ref9}. Previous work by \cite{ref8} achieved FSIM 0.920 on fast spin-echo using densely connected U-Nets with GAN-guided training and perceptual loss. Tamada et al. \cite{ref9} used an 8-layer U-Net for abdominal motion artifact reduction. Kromrey et al. \cite{ref10} extended this to multi-arterial phase liver imaging using multi-channel CNNs. These studies were conducted on 3.0 T or 1.5 T systems in East Asia and Europe, with no validation on African cohorts or low-field dominated workflows.
\subsubsection{2.2 K-space and dual-domain methods: the raw-data barrier:} 
MoPED estimates rigid-body motion from Cartesian k-space center patches embedded in iterative data-consistency algorithms, reducing computation 20-fold \cite{ref11}. D2MCNet combines k-space uncertainty modules with image-domain reconstruction \cite{ref12}. Both require raw k-space, multi-coil sensitivity profiles, and vendor research agreements inaccessible outside academic centers \cite{ref11,ref12}. The CUPID framework \cite{ref13} bypasses raw-data barriers for reconstruction using only DICOM images, but for motion correction specifically, image-domain networks remain the only viable retrospective solution in most LMIC centers.
\subsubsection{2.3 Synthetic motion simulation and domain generalisation:}
Synthetic simulation is standard when real pairs are scarce. Early work added linear phase errors in k-space \cite{ref17}; modern approaches employ 3D elastic displacement fields and signal-loss models \cite{ref8}. Domain shift remains central: intensity and contrast shifts between sequences cause catastrophic failure unless regularized \cite{ref15,ref16}. Previous studies proposed deep stacked transformations (BigAug) to simulate cross-scanner variations, reducing Dice degradation from 39 \% to 11 \%. In our setting, synthetic motion generation is a necessity imposed by PACS deletion of failed scans, not a methodological shortcut \cite{ref13,ref14}.
\section{Methodology}
\subsection{Study setting and dataset}
This retrospective study was conducted at a center that operates 1.5 T Siemens Magnetom Essenza scanners without vendor motion-correction licenses. After screening 847 abdominal MRI studies (January 2022-March 2026), inclusion criteria were: axial T2 HASTE, slice thickness 2.5-8.5 mm, TE 80-150 ms, sufficient anatomical coverage, absence of gross truncation or metal artifacts. Exclusions: post-surgical status, ascites obscuring boundaries and incomplete DICOM series. 
This yielded 139 clean HASTE (clean cohort), 25 HASTE with visible motion artifacts identified by a board-certified radiologist (A.R.M., 8 years) (real-motion cohort), and 79 patients with paired HASTE + free-breathing TSE-TRIGGER (paired cross-sequence cohort). (Table~\ref{tab1}) summarises acquisition parameters.

\subsection{Ethics and anonymisation}
Approved by National Health Research Ethics Committee Board, 
Informed consent was waived for retrospective de-identified data. Anonymisation: (i) DICOM metadata scrubbing via \texttt{pydicom}; (ii) 20-pixel edge cropping to remove burned-in text; (iii) filename replacement with non-identifiable codes. A secure institutional JSON mapping file is maintained for audit purposes.
\subsection{Image preprocessing}
DICOM converted to NIfTI via dcm2niix (v6.0.0), preserving 256 × 256 matrix. Volumes resampled to isotropic 1.0 mm slice direction. Intensity normalised to [0, 1] via min-max scaling with 3.5 SD outlier clipping. Network receives full abdominal field of view. Clean volumes serialised as .npz with key data.

\subsection{Synthetic respiratory motion generation and data split}

Because severely degraded acquisitions are deleted during PACS cleanup, real failed breath-holds were unavailable for supervised training. Synthetic motion was a necessity imposed by infrastructure, not convenience. We implemented v8: 3D elastic warp (Gaussian $\sigma = 2$ in-plane, $\sigma = 4$ through-plane, $\alpha = 6$ moderate / $4$ mild), modulated by diaphragm-dome model (maximum displacement at lung--liver interface, tapering to 20\% at lower abdomen). Ghost replicas used blend physics (not additive): shifted copy blended with weight $w = 0.18$--$0.22$, preventing intensity clipping. Rigid shift ($\pm 2$ pixels), edge signal loss (top 4\% gradient pixels attenuated 6--8\%), Gaussian blur ($\sigma = 0.25$--$0.35$), and Rician noise ($\sigma = 0.005$--$0.007$). Severity: 40\% mild, 60\% moderate; severe motion excluded. The dataset was split to 139 patients shuffled (seed = 42): 100 train, 20 validation, 19 test. Mean max diff: $0.319 \pm 0.082$ (range $0.110$--$0.580$); 18\% $> 0.50$.
\subsection{Model architecture}
Standard 2D U-Net with skip connections \cite{ref7}. Encoder: $64 \rightarrow 128 \rightarrow 256 \rightarrow 512$ filters, each with two $3 \times 3$ convolutions + BatchNorm + ReLU, followed by $2 \times 2$ max-pooling. Bottleneck: 512 filters. Decoder: transposed convolutions upsampling, concatenating encoder skip connections, with matching convolutional blocks. Final $1 \times 1$ convolution outputs single-channel $256 \times 256$ grayscale. Total parameters: $\sim 7.7 \times 10^6$. 2D slice-wise chosen because: (i) 139 patients insufficient for 3D U-Net without overfitting; (ii) limited GPU memory; (iii) through-plane motion minimal in breath-hold HASTE.
Loss: L1 (mean absolute error). L1 chosen over L2 for sharper edges and stability on limited data.
\subsection{Training protocol and compute resources}
Google Colab, NVIDIA Tesla T4 GPU (16 GB VRAM), Intel Xeon CPU (2 vCores), 12.7 GB RAM. Python 3.12, PyTorch 2.1.0, CUDA 12.1.
Dataset preloaded into RAM. Training: 30 random slices/patient/epoch = 3,000 samples/epoch, ~120,000 over 40 epochs. No augmentation; random slice sampling sufficient.
\[
\text{Adam: } \beta_1 = 0.9,\; \beta_2 = 0.999,\; \epsilon = 1 \times 10^{-8};\quad \text{lr} = 1 \times 10^{-4}
\]
batch size 8, mixed precision AMP, GradScaler. Early stopping: patience 20 epochs monitoring validation L1. Training: $\sim$73 s/epoch, total $\sim$2.5 hours. Inference: 45 ms/slice GPU, 1.35 s/volume GPU, 2.8 s/volume CPU.
\subsection{Evaluation strategy (three-tier)}
Tier 1 (Engineering): 19 held-out synthetic patients. SSIM (Gaussian window $\sigma = 1.5$, size $11 \times 11$, dynamic range 1.0) and \[
\text{PSNR} = 20 \log_{10}\left(\frac{1.0}{\sqrt{\text{MSE}}}\right)
\] computed slice-wise, averaged per volume.
Tier 2 (Clinical same-sequence): 25 real motion HASTE cases processed without retraining. Blinded Likert scoring (1 = non-diagnostic, 2 = poor, 3 = acceptable, 4 = good, 5 = excellent) by board-certified radiologist and senior MRI radiographer (8 and 10 years experience) respectively, independently, on central 10 slices. Raters blinded to condition and to each other.
Tier 3 (Cross-sequence): 79 paired TRIGGER applied to HASTE-trained model without retraining. SSIM/PSNR vs paired HASTE pseudo-ground-truth. Blinded Likert scoring on 20 random TRIGGER cases.
SSIM and PSNR: mean $\pm$ SD. Likert comparisons: Wilcoxon signed-rank test (two-sided $P < 0.05$). Inter-rater reliability: weighted Cohen's $\kappa$ and ICC.

\section{Results}
On the 19 held-out synthetic test patients, the U-Net achieved a mean structural similarity index measure (SSIM) of 0.863 ± 0.042 (range 0.731–0.926), a mean peak signal-to-noise ratio (PSNR) of 30.06 ± 1.80 dB (range 26.22–33.15), and a mean absolute error of 0.016 ± 0.003 against the known clean targets; the baseline SSIM between the synthetic motion input and the clean target was 0.816 ± 0.059. For the 25 native motion-degraded HASTE cases. Inter-rater reliability was weighted Cohen's = 0.644 and ICC = 0.712 for original images, and = 0.535 and ICC = 0.646 for corrected images; exact agreement was 80.0 \% and 68.0 \% respectively, with 100 \% agreement within one Likert point for both conditions. Mean Likert scores across both raters were 4.06 ± 0.55 for original images and 4.04 ± 0.61 for corrected images (Wilcoxon signed-rank W = 44.0, P = 0.914); six cases (24 \%) improved, twelve (48 \%) were unchanged, and seven (28 \%) were worse after correction. For the 79 paired TSE-TRIGGER acquisitions, the HASTE-trained model was applied without retraining and evaluated against the paired clean HASTE as pseudo-reference. Corrected TRIGGER achieved SSIM 0.404 ± 0.073 versus 0.334 ± 0.055 for uncorrected TRIGGER (paired t = 26.34, P < 0.001) and PSNR 14.92 ± 2.04 dB versus 14.85 ± 1.96 dB (t = 7.17, P < 0.001). Blinded Likert scoring of 20 randomly selected TRIGGER cases yielded a median score of 3.0 (IQR 2.0-3.0) for both uncorrected and corrected conditions (P = 0.38). A summary of the three-tier validation is provided in Table 2.

\begin{table}[H]
\centering
\caption{Three-tier validation results for the NIMARC-MRI motion correction pipeline. Likert scale 1--5 (1 = non-diagnostic, 5 = excellent).}\label{tab2}
\begin{tabular}{|l|c|c|c|c|c|}
\hline
Cohort & $n$ & SSIM & PSNR (dB) & \makecell{Likert mean \\ $\pm$ SD} & $P$ value \\
\hline
Tier 1 (Synthetic test) & 19 & $0.863 \pm 0.042$ & $30.06 \pm 1.80$ & --- & --- \\
Tier 2 (Real motion HASTE) & 25 & --- & --- & \makecell{$4.06 \pm 0.55$ \\ vs $4.04 \pm 0.61$} & $0.914$ \\
Tier 3 (Paired TRIGGER) & 79 & \makecell{$0.404 \pm 0.073$ \\ vs $0.334 \pm 0.055$} & \makecell{$14.92 \pm 2.04$ \\ vs $14.85 \pm 1.96$} & --- & $<0.001$ \\
\hline
\end{tabular}
\end{table}

\begin{table}\centering
\caption{MRI acquisition parameters for the NIMARC-MRI dataset sequences.}\label{tab1}
\begin{tabular}{|c|c|c|}
\hline
\makecell[c]{Parameter} & \makecell[c]{T2 HASTE \\ (Clean \& Motion)} & \makecell[c]{T2 TSE-TRIGGER \\ (Paired)} \\
\hline
\makecell[c]{Scanner} & \makecell[c]{Siemens Magnetom \\ Essenza 1.5 T} & \makecell[c]{Siemens Magnetom \\ Essenza 1.5 T} \\
\makecell[c]{Sequence} & \makecell[c]{Single-shot \\ turbo spin-echo} & \makecell[c]{Multi-shot \\ respiratory-triggered TSE} \\
\makecell[c]{Plane} & Axial & Axial \\
\makecell[c]{TR (ms)} & 826 $\pm$ 120 & 4,574 $\pm$ 890 \\
\makecell[c]{TE (ms)} & 80--150 & 85--120 \\
\makecell[c]{Echo train \\ length} & 77--96 & 12--26 (mean 19) \\
\makecell[c]{Slice thickness \\ (mm)} & 2.5--8.5 & 4.0--6.0 \\
\makecell[c]{Matrix} & 256 $\times$ 256 & 256 $\times$ 256 \\
\makecell[c]{Fat \\ suppression} & SPAIR / None & SPAIR \\
\makecell[c]{Breathing} & \makecell[c]{Breath-hold \\ (single-shot)} & \makecell[c]{Free-breathing, \\ belt triggering} \\
\makecell[c]{Acquisition \\ time} & $\sim$18--22 s per slab & $\sim$3--5 min per slab \\
\hline
\end{tabular}
\end{table}

\begin{figure}
\centering
\includegraphics[width=0.90\textwidth]{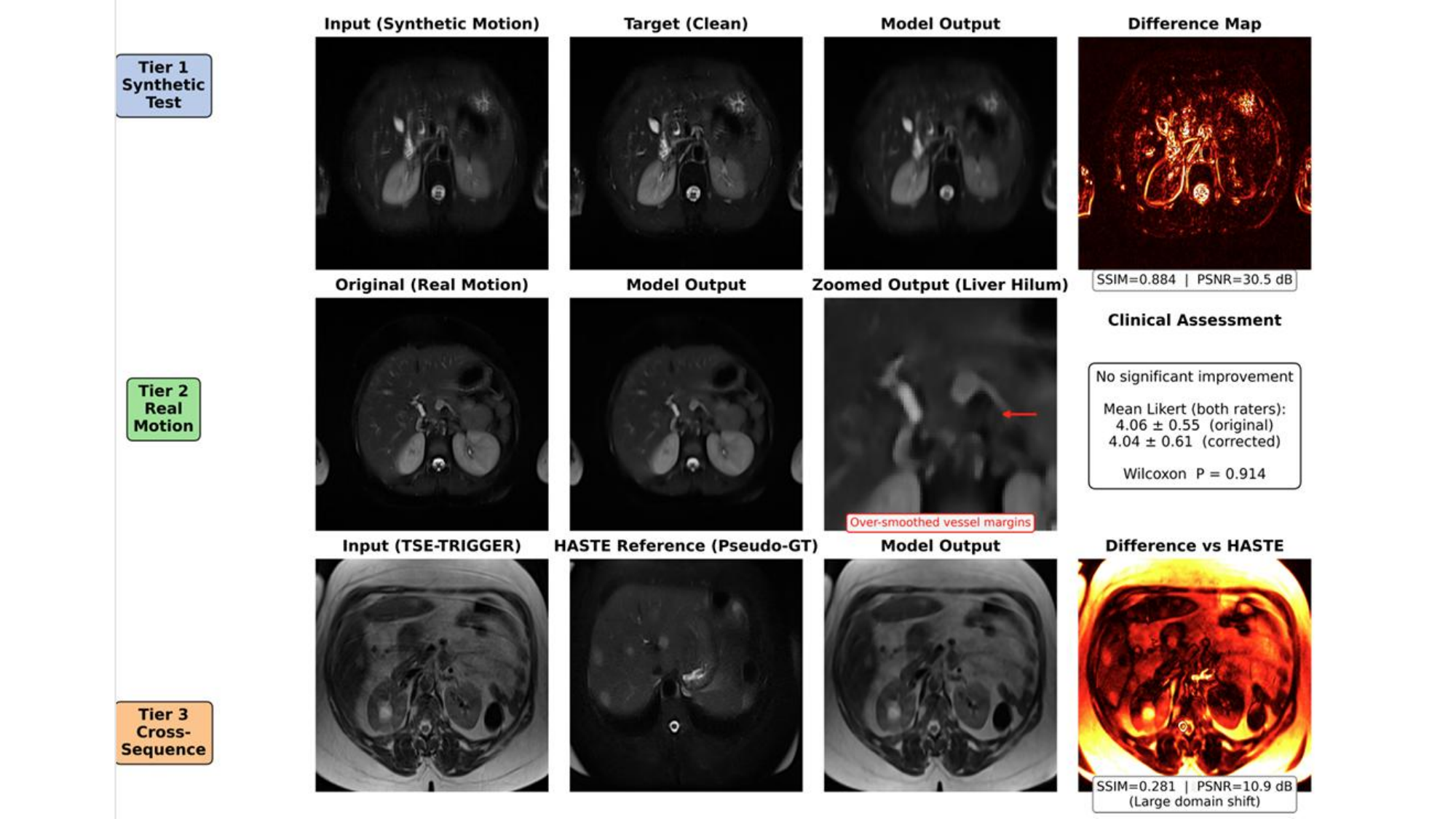}

\caption{Three-tier validation of the HASTE-trained U-Net on NIMARC-MRI. \textbf{Row 1 (Tier 1 -- Synthetic test):} The model successfully inverts synthetic motion artifacts (SSIM 0.863, PSNR 30.06 dB). \textbf{Row 2 (Tier 2 -- Real clinical motion):} No perceptual improvement on native motion-degraded HASTE (mean Likert 4.06 vs 4.04, $P = 0.91$); red arrow indicates over-smoothed hepatic vessel margins. \textbf{Row 3 (Tier 3 -- Cross-sequence TRIGGER):} Application to a fundamentally different acquisition protocol yields marginal SSIM improvement (0.404 vs 0.334) but no perceptual gain, with large residual difference against the HASTE reference reflecting domain shift in T2 weighting and SNR.}
\label{fig1}
\end{figure}

\section{Discussion}
Our model achieved high quantitative fidelity on synthetic test data (Table~\ref{tab2}), but failed to improve native clinical motion artifacts. The real-motion cases were already rated ``Good'' and corrected images showed no significant improvement. This synthetic-to-real gap arises because conservative synthetic motion (mean max diff 0.319) did not encompass clinically relevant severity, and because L1 loss encourages blurry, conservative outputs that suppress high-frequency texture~\cite{ref8}. Inter-rater agreement dropped from substantial ($\kappa_w = 0.644$) on original images to moderate ($\kappa_w = 0.535$) on corrected images, suggesting the model introduced ambiguous artifacts that created divergent perceptual assessments.

A critical insight is that T2 HASTE remains the clinical default because its $\sim$18--22~s breath-hold acquisition sustains high patient throughput and delivers superior spatial resolution (2.5--8.5~mm slices, ETL 77--96) essential for small-lesion detection, whereas TSE-TRIGGER's 3--5~min free-breathing acquisition creates a severe throughput bottleneck with thicker slices and altered T2 weighting that degrades diagnostic confidence; it is reserved only for non-cooperative patients, making vendor-neutral HASTE rescue a genuine clinical necessity rather than a sequence preference. Clinical workflow inherently pre-filters severely degraded scans: technologists repeat or reject non-diagnostic breath-holds at the scanner console, and severely degraded acquisitions are subsequently deleted during routine PACS cleanup to manage storage constraints an institutional reality well-documented in LMIC radiology ecosystems \cite{ref14}. The 25 real-motion cases that survived this pre-filtering were therefore predominantly mild-to-moderate (Likert 3--4), leaving the model no severely degraded inputs to improve. This aligns with the broader finding that synthetic-to-real transfer in MRI motion correction is fundamentally limited when training simulations do not span the full clinical severity distribution \cite{ref8,ref18}.

Future work must advance beyond fixed-parameter synthetic augmentation toward severity-adaptive simulation calibrated against clinical statistics via deep stacked transformations, domain-randomisation strategies modeling cross-scanner and cross-protocol variation \cite{ref15,ref17}, or unsupervised adversarial and motion-estimation frameworks that eliminate the need for paired clean targets \cite{ref19,ref20} while the cross-sequence TRIGGER evaluation (Table~\ref{tab2}) establishes the sequence generalization limit of the HASTE-trained model: although structural similarity improves modestly through transferable smoothing and noise suppression, negligible gains in peak signal-to-noise ratio and unchanged blinded Likert scores confirm that image-domain networks cannot recover TRIGGER-specific contrast \cite{ref12,ref16}, as the domain shift between single-shot breath-hold HASTE and respiratory-gated multi-shot TRIGGER exceeds post-processing capacity. 

Several limitations must be acknowledged. The cohort is small ($n=139$) relative to Western datasets and derives from a single center and scanner model. The 2D slice-wise architecture ignores through-plane motion and inter-slice coherence, while magnitude-only processing discards phase information that could distinguish motion ghosts from true anatomy \cite{ref11,ref12}. The real-motion subset ($n=25$) was identified by a single reader, limiting statistical power and introducing selection bias; the 5-point Likert scale further exhibited a ceiling effect (mean $\sim$4.0), reducing sensitivity to subtle changes, and we report inter-rater reliability descriptively without formally testing the post-correction drop in agreement. No loss-function ablation was performed, so the contribution of L1-induced blurring cannot be isolated. Cross-sequence metrics confound motion correction with inherent contrast and SNR differences between HASTE and TSE-TRIGGER. Finally, downstream diagnostic performance was not evaluated due to the absence of histopathological and follow-up confirmation. Nevertheless, the enduring contribution is NIMARC-MRI dataset. Future work will explore severity-adaptive simulation, perceptual and adversarial losses, fine-tuning on larger real-motion cohorts, and benchmarking against traditional motion-estimation baselines.

\section{Impact in RCS}

Abdominal MRI research in resource-constrained settings is paralyzed by two barriers: vendor research agreements block raw k-space access, and routine PACS cleanup deletes failed scans, erasing the paired data required for supervised training. With no public abdominal MRI datasets from West Africa, LMIC sites lack both benchmarks and vendor-neutral alternatives. This study introduces NIMARC-MRI, the first public abdominal MRI dataset from the region, comprising clean, motion-degraded, and cross-sequence acquisitions from a Nigerian 1.5 T center without motion-correction licenses. NIMARC-MRI is publicly available on Zenodo (\url{https://doi.org/10.5281/zenodo.22115936}) under a controlled data-use agreement. By training a 2D U-Net entirely on reconstructed DICOM images and validating it through a three-tier strategy, we establish a reproducible baseline that requires no raw data access.

\begin{credits}
\subsubsection{\ackname}
The authors thank the following for their support: the Lacuna Fund for Health and Equity (PI: Udunna Anazodo, 0508-S-001); the Natural Sciences and Engineering Research Council of Canada (NSERC) Discovery Launch Supplement (PI: Udunna Anazodo, DGECR-2022-00136); the Digital Research Alliance of Canada; the McGill University Doctoral Internship Program; the University of Washington Azure GenAI for Science Hub; the SPARK Academy 2025 instructors; the African Institute for Research Advancement and Innovation (AIRA Africa); and Medserve Kano Diagnostic Center for their tremendous support and funding.
\subsubsection{\discintname}
The authors have no competing interests to declare that are relevant to the content of this article.
\end{credits}

\end{document}